\documentclass[aps,prb,superscriptaddress,floatfix,twocolumn]{revtex4-2}

\usepackage{bm}        
\usepackage{amsmath,amssymb,amsthm}   
\usepackage{textgreek}
\usepackage{mathtools}
\usepackage{tabularx}
\usepackage{array}
\usepackage{color,soul}
\usepackage{diagbox}
\usepackage{xcolor}
\usepackage{natbib}
\usepackage{siunitx} 
\usepackage{hyperref}

\usepackage{grffile}
\usepackage{graphicx}
\usepackage{float}
\usepackage{chemformula}
\usepackage{gensymb}
\usepackage{multirow}

\definecolor{mygreen}{rgb}{0.1,0.6,.1}
\definecolor{myred}{rgb}{.6,.1,.1}
\definecolor{myblue}{rgb}{.1,.1,.8}
\definecolor{mycyan}{rgb}{0,.8,.8}
\definecolor{mymagenta}{rgb}{0.9,0,0.9}

\newcommand{\wn}{\,cm$^{-1}$}
\newcommand{\Ag}[1]{A$_{g}^{#1}$}
\newcommand{\Bg}[1]{B$_{g}^{#1}$}

\newcommand{\TN}{{$T_\textrm{N}$}}

\newcommand{\Au}[1]{A$_{u}^{#1}$}
\newcommand{\Bu}[1]{B$_{u}^{#1}$}

\newcommand{\Ph}[1]{P$_{#1}$}
\newcommand{\M}[1]{M$_{#1}$}
\newcommand{\degr}{$^\circ$}

\begin{document}
\title{Spin-lattice entanglement in \ch{CoPS_3}}
\author{Thuc T. Mai}%
\email{thuc.mai.ctr@us.af.mil}
\affiliation{Quantum Measurement Division, Physical Measurement Laboratory, NIST, Gaithersburg, MD 20899}
\affiliation{Materials and Manufacturing Directorate, Air Force Research Laboratory, Wright-Patterson Air Force Base, OH}
\affiliation{Bluehalo, LLC, 4401 Dayton-Xenia Rd, Dayton, 45432, OH, USA}
\author{Amber McCreary}%
\affiliation{Quantum Measurement Division, Physical Measurement Laboratory, NIST, Gaithersburg, MD 20899}
\author{K.F. Garrity}%
\affiliation{Materials Measurement Science Division, Materials Measurement Laboratory, NIST, Gaithersburg, MD 20899}
\author{Rebecca L. Dally}
\affiliation{Sensor Science Division, Physical Measurement Laboratory, National Institute of Standards and Technology, Gaithersburg, MD 20899}
\author{Sambridhi Shah}
\affiliation{Department of Chemistry, University of Tennessee, Knoxville, TN 37996, USA}
\author{Bryan C. Chakoumakos}%
\affiliation{Neutron Scattering Division, Oak Ridge National Laboratory, Oak Ridge, TN 37830}
\author{Md Nasim Afroj Taj}
\affiliation{Electrical and Computer Engineering Department,  University of Virginia, Charlottesville,  VA 22904}
\author{Jeffrey W. Lynn}
\affiliation{NIST Center for Neutron Research, National Institute of Standards and Technology, Gaithersburg, MD 20899}
\author{Michael A. McGuire}%
\affiliation{Materials Science and Technology Division, Oak Ridge National Laboratory, Oak Ridge, TN 37831}
\author{Benjamin S. Conner}
\affiliation{Materials and Manufacturing Directorate, Air Force Research Laboratory, Wright-Patterson Air Force Base, OH}
\affiliation{Azimuth Corporation, 2079 Presidential Dr. No. 200, Fairborn, OH 45342, USA}
\author{Mona Zebarjadi}
\affiliation{Electrical and Computer Engineering Department,  Materials Science and Engineering Department,  University of Virginia, Charlottesville,  VA 22904}
\author{Janice L. Musfeldt}
\affiliation{Department of Chemistry, University of Tennessee, Knoxville, TN 37996, USA}
\affiliation{Department of Physics and Astronomy, University of Tennessee, Knoxville, TN 37996, USA}
\author{Angela R. Hight Walker}
\affiliation{Quantum Measurement Division, Physical Measurement Laboratory, NIST, Gaithersburg, MD 20899}
\author{Rahul Rao}
\affiliation{Materials and Manufacturing Directorate, Air Force Research Laboratory, Wright-Patterson Air Force Base, OH}
\author{Michael A. Susner}
\email{michael.susner.2@us.af.mil}
\affiliation{Materials and Manufacturing Directorate, Air Force Research Laboratory, Wright-Patterson Air Force Base, OH}

\begin{abstract}

Complex chalcogenides in the $M$PS$_3$ family of materials ($M$ = Mn, Fe, Co, and Ni) display remarkably different phase progressions depending upon the metal center orbital filling, character of the P–P linkage, and size of the van der Waals gap. There is also a stacking pattern and spin state difference between the “lighter” and “heavier” transition metal-containing systems that places CoPS$_3$ at the nexus of these activities. Despite these unique properties, this compound is under-explored. Here, we bring together Raman scattering spectroscopy and infrared absorption spectroscopy with X-ray techniques to identify a structural component to the 119 K magnetic ordering transition as well as a remarkable lower temperature set of magnon-phonon pairs that engage in avoided crossings along with a magnetic scattering continuum that correlates with phonon lifetime effects. These findings point to strong spin-phonon entanglement as well as opportunities to control these effects under external stimuli.

\end{abstract}
\maketitle

\section{Introduction}

Van der Waals (vdW) solids are superb platforms for exploring the interplay between structure and magnetism in reduced dimensions \cite{mak_probing_2019}. Complex chalcogenides in the \ch{MPX_3} family  \cite{Susner_2017} are especially interesting because they host extensive chemical and physical tunability. Here, M is a transition metal, P is phosphorus, and X is either sulfur or selenium. For M =  Mn, Fe, Co, and Ni, the system hosts a variety of antiferromagnetic (AFM) arrangements  ranging from zig-zag Ising AFM order in \ch{FePX_3} to in-plane zig-zag AFM order in \ch{CoPS_3} and \ch{NiPS_3} to easy-plane N\'eel-type AFM order in \ch{MnPSe_3}  \cite{Susner_2017}. At the single ion level, the 2+ charge on the metal ion results in a rich many-body electronic ground state.  Examples include a pure spin configuration of S=5/2 for Mn$^{2+}$ and mixed spin-orbit coupled states including S=1, L=3 for Ni$^{2+}$, S=2, L = 1 for Fe$^{2+}$, and  S=3/2,  L=3  for Co$^{2+}$ \cite{AandBbook}. 
Of all the \ch{MPX_3} compounds, \ch{CoPS_3} is one of the least studied members of this family of materials, primarily due to challenges with synthesis. It hosts AFM order below \TN\:$\approx$ 119\,K \cite{wildes_CoPS3_2017}. At the same time,  spin-orbit coupling emanating from the Co$^{2+}$ center is highly under-explored, although based upon recent measurements of other honeycomb lattice magnets like \ch{CoTiO_3} \cite{elliot_CoTiO3_2021,Li_CoTiO3_2024,Mai_2025}, the interaction is likely to impart additional functionality in the form of a rich magnetic excitation spectrum that can be described by both spin and orbital degrees of freedom. We therefore anticipate that spin-orbit coupling (SOC) will entangle the magnetic and crystalline structures in CoPS$_3$ in a meaningful and potentially useful manner. 

Perhaps unsurprisingly, observations of spin-lattice coupling across the AFM transition temperatures have been reported for several members of the \ch{MPX_3} family, although only \ch{FePS_3} has been confirmed to exhibit a concomitant structural transition with its magnetic transition through diffraction \cite{Bjarman1983,JERNBERG1984} and calorimetric measurements \cite{TAKANO2004}. Spectroscopic measurements on bulk crystals as well as mono- and few-layered  \ch{MPX_3} materials have shown the appearance of new zone-folded phonon modes\cite{lee_FePS3_2016,McCreary_FePS3_2020,Sun_2024}, increased two-magnon scattering and Fano resonances below \TN\:  \cite{Rosenblum_NiPS3_1994,kim_NiPS3_2019}, as well as strong variations in peak frequencies and intensities due to spin-lattice coupling \cite{lee_FePS3_2016,mai_MnPSe3_2021,Rao_2024}. The hybridization or entanglement between magnons and phonons below  \TN\:was also reported for \ch{MPX_3} compounds\cite{mai_MnPSe3_2021,Liu_FePS3_2021,cui_FePSe3_2023,luo_FePSe3_2023,lemardele_FePSe3_2024}. Here, magnons and phonons are the quasiparticles that correspond to the magnetic and lattice excitations, respectively. 

An inelastic neutron scattering study on \ch{CoPS_3} single crystals revealed four magnon branches at the Brillouin Zone center, with the lowest two being approximately 15 meV and 23 meV\cite{wildes_CoPS3_2023}. While significant overlap in the magnon and phonon energies was also reported, spin-lattice interactions were not considered. In this study, we uncovered two key and unambiguous signatures of spin-lattice coupling: magnetostriction and magnon-phonon hybridization.

We conducted detailed temperature-dependent Raman spectroscopy, IR spectroscopy, single crystal X-ray diffraction (XRD), and high-resolution synchrotron powder XRD to characterize the lattice and magnetic excitation spectra of single-crystalline \ch{CoPS_3}. The temperature-dependent powder XRD shows a subtle lattice contraction concomitant with the magnetic transition, which was further confirmed in our specific heat measurements, highlighting the strong spin-lattice coupling in \ch{CoPS_3}. We also observed two sharp peaks at 106.9\wn\: and 187.5\wn\:(13.3 meV and 23.2 meV, respectively) in the low temperature (5 K) Raman data, corresponding to the two lowest zone-center magnons. The temperature-dependent Raman data reveal several unique features that can be explained by the hybridization of the magnons and the Raman-active phonons. Our study reveals in detail the specific interactions between lattice excitations and magnetic excitations, paving the way for future studies of the rich physics that awaits us at the 2D limit of these layered magnetic materials.

\section{Methods}

We synthesized single crystals of \ch{CoPS_3} using the general procedures given in \cite{Koughnet2024,Matsuoka2023,Susner_2017}. We combined, in a near stoichiometric ratio (20\% excess P) freshly reduced Co powder (Alfa Aesar Puratronic, 22 mesh, 99.998\%), P chunks (Alfa Aesar Puratronic, 99.999\%), and S chunks (Alfa Aesar Puratronic, 99.9995\%) in an evacuated fused silica ampoule (20 cm length, 14 mm ID, 19 mm OD)  together with $\approx$ 100 mg \ch{I_2} as a vapor transport agent. We then placed the ampoule into an MHI H-series tube furnace, ramped to the reaction temperature of 650 \degr C over 30 h, and held the reaction at that temperature for 100 h, after which we cooled to room temperature over an additional period of 30 h. The crystals we obtained were in the form of micaceous, flat platelets $<$ 0.5 mm thick with the c axis aligned with the plane normal of the platelet. The largest crystals were $\sim$ 20 mm in diameter with most being in the 4 mm to 7 mm range.

Next, we confirmed the composition with electron microscopy–energy-dispersive X-ray spectroscopy (SEM-EDS) analysis (3–4 spots per crystal for 9–12 spots total per batch) using a Thermo Scientific UltraDry EDS spectrometer joined with a JEOL JSM-6060 SEM. These results came out to, within error, the composition \ch{CoPS_3}. We employed a Quantum Design MPMS-3 superconducting quantum interference device magnetometer to take magnetization measurements. Additionally, we collected XRD spectra off a flat crystal face (00l) using a Philips PANalytical system employing Cu K$_\alpha$ X-ray radiation at a wavelength of $\lambda_{\textrm{K}_{\alpha1}}$ = 0.154056 nm. We performed refinements on these data using FullProf \cite{FRONTERA2004} with Le-Bail fits to elucidate the layer spacing (here the monoclinic angle was set to 90\degr). We used a Quantum Design Dynacool PPMS instrument to measure specific heat capacity data via pulsed calorimetry with a 2\% heat rise. We also used the dual-slope analysis option with a 10\% and 20\% heat rise to probe possible phase transitions below \TN.

We performed single crystal X-ray diffraction on a Rigaku XtalLAB Synergy S diffractometer equipped with a kappa-goniometer by using a Mo X-ray source. The data collection and reduction procedures were conducted in the CrysAlis$^\textrm{Pro}$ software\cite{Crysalispro}. High-resolution powder X-ray diffraction data were taken at the Advanced Photon Source synchrotron at Argonne National Laboratory on the beamline 11-BM. Single crystals of \ch{CoPS_3} were ground into a powder and encapsulated in a Kapton capillary which was placed into a closed-cycle cryostat reaching a base sample temperature of 6 K. Data were taken using a wavelength of $\lambda$ = 0.458122 \textup{\AA}. The diffraction data were analyzed using a LeBail fit and the FullProf software \cite{FRONTERA2004}, where the fit used the reflection conditions for the C2/m space group.

For the temperature-dependent Raman measurements, we mounted a freshly cleaved bulk crystal  inside a close-cycled optical cryostat in a helium gas environment. The sample temperature could be varied from 2K to 300K. The 633 nm excitation beam from a HeNe laser was incident onto the sample via free-space optics, through a 50X cryogenic objective. The laser spot size was approximately 1 \textmu m. The back-scattered beam was then collected and sent to a triple-stage spectrometer operating in triple-subtractive mode, which rejects the elastic light (Rayleigh scattering) and sends the Raman scattering to a liquid nitrogen cooled CCD. Various polarization optics were used to control the incident polarization and the scattered polarization. VV (VH) denotes the incident and scattered polarization being parallel (perpendicular) to each other. Polarization-resolved spectra were collected by placing a half-wave plate in the common beam path of the incident and scatter light, thus rotating both polarizations. This is equivalent to physically rotating the sample \cite{Mai_RuCl3_2019}. 

A single crystal of CoPS$_3$ was exfoliated between polypropyene tape to create a sample with appropriate optical density for our infrared (IR) spectroscopy experiments. We measured the far IR transmittance using a Bruker 113v equipped with a helium-cooled bolometer detector and converted the transmittance to absorption. A continuous flow cryostat provided temperature control. 

We performed first principles density functional theory (DFT)\cite{hk,ks,phonopy-phono3py-JPSJ} calculations using the Quantum Espresso code\cite{qe}, with GBRV pseudopotentials\cite{gbrv}. We used the PBEsol exchange correlation function functional\cite{pbesol} and a Hubbard U correction of 4 eV on the Co-d states (DFT+U)\cite{DFTU}. We used the phonopy code to calculate phonon frequencies and perform symmetry analysis \cite{phonopy-phono3py-JPCM,phonopy-phono3py-JPSJ}.

\section{Results}

\begin{figure*}[]
\includegraphics[width=0.75\textwidth]{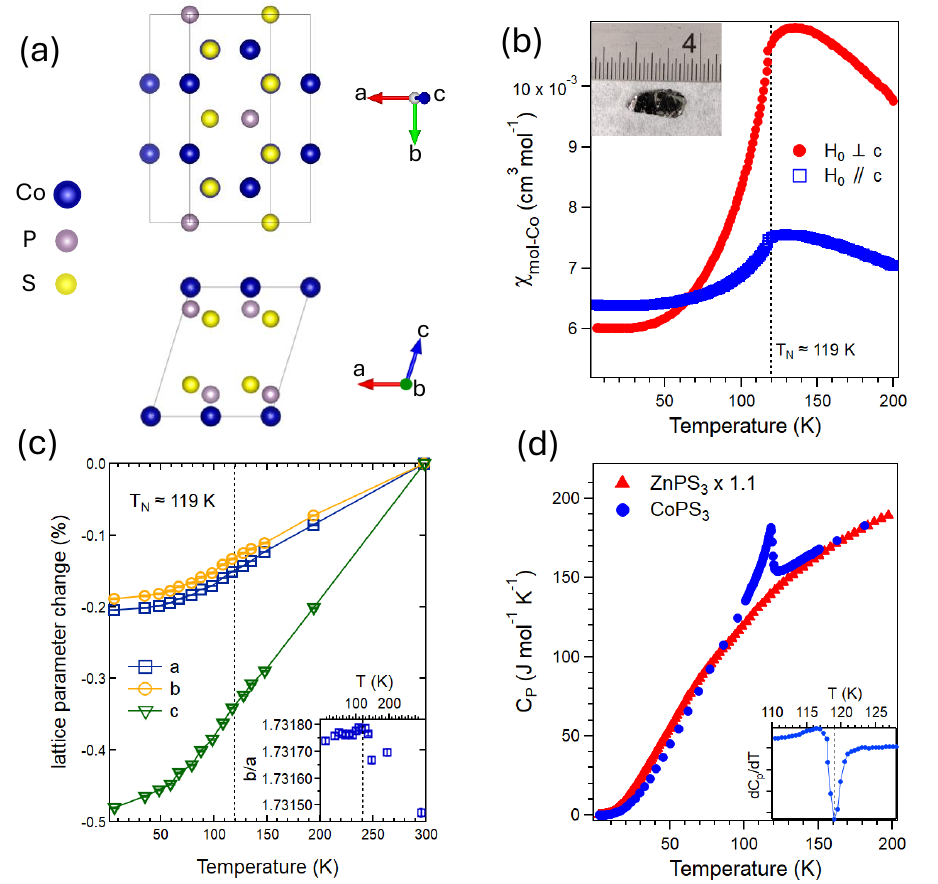}
\caption{\textbf{Characterization of CoPS$_3$ single crystals}. (a) Crystal structure views of the a-b plane (top) and a-c plane (bottom), with the monoclinic unit cell outlined. The images were generated with VESTA\cite{VESTA}. (b) The magnetic susceptibility (M/H) as a function of temperature with \TN\:marking the onset of antiferromagnetic ordering. The inset shows picture of a large single crystal. (c) From high-resolution synchrotron X-ray powder diffraction, the change in lattice parameters a, b, and c are shown across \TN. The inset shows the ratio of b/a, where the ratio freezes below \TN. (d) Specific heat of \ch{CoPS_3} showing a sharp peak also at \TN, with an inset showing the derivative of the heat capacity curve. For comparison, the heat capacity of non-magnetic \ch{ZnPS_3} is also shown. For both materials, 10 atoms per unit cell were used to obtain C$_\mathrm{P}$. }
\label{fig:characterization}
\end{figure*}
The crystallographic unit cell of \ch{CoPS_3} is shown in Figure \ref{fig:characterization}(a), where the monoclinic stacking along the c-direction breaks the 3-fold rotation symmetry of the Co$^{2+}$ honeycomb pattern in a single vdW layer. The inset of Figure \ref{fig:characterization}(b) shows a picture of one of our \ch{CoPS_3} single crystals, which typically grow to several mm in size. The temperature-dependent magnetic susceptibility of \ch{CoPS_3} points to an in-plane antiferromagnet, with the transition temperature being approximately 119 K (Fig. \ref{fig:characterization}(b)), which is consistent with the literature \cite{wildes_CoPS3_2017}. The reported ordered moment is slightly greater than 3 $\mu_B$, more than the spin-only contribution from the Co$^{2+}$ ions \cite{wildes_CoPS3_2017}, signifying an orbital component to the magnetism. Refined neutron diffraction analyzed in the critical regime from 109 K to \TN\: revealed the critical exponent for the staggered magnetization is \textbeta = 0.30(1), indicating that the order is more three dimensional than some of these vdW materials, but is well below the value for the isotropic Heisenberg model of \textbeta = 0.365 \cite{wildes_CoPS3_2017}.

The room temperature XRD pattern from an 00\textit{L} face of a single crystal of \ch{CoPS_3} exhibits narrow linewidths, confirming the high crystal quality of our material (Fig. S1). In addition, our initial single crystal XRD measurements confirmed the C2/m space group at three different temperatures: 95 K, 150 K, and 250 K, consistent with previously published studies \cite{OUVRARD_1985,wildes_CoPS3_2017} (See SI section I). We emphasize that the bulk \ch{CoPS_3} crystal lattice is monoclinic (point group C$_{\rm{2h}}$), not hexagonal (D$_{\rm{3d}}$), at room temperature and exhibits no evidence of \textit{additional} rotational symmetry breaking across \TN. In order to reveal the subtle temperature-dependent changes in lattice parameters, we performed a higher-resolution study using a synchrotron source on a powder sample. In Fig. \ref{fig:characterization}(c), we show the lattice parameters obtained from a LeBail fit of the high-resolution synchrotron data with respect to temperature using the reflection conditions for the C2/m space group. The lattice parameters in Fig. \ref{fig:characterization}(c) are expressed in percentage of their values at 298 K, where a = 5.89901(3) \textup{\AA}, b = 10.21398(9) \textup{\AA}, c = 6.66621(6) \textup{\AA}, and $\mathrm{\beta}$ = 107.1938(6)\degr\:(uncertainties represent one standard deviation). As expected for a van der Waals material, the stacking direction \textit{c} exhibits the largest change as the lamellae expand and contract, accordion like, with changing temperature. However, we note that, near \TN\: there is a subtle change in slope of the temperature dependence, depicted by the dashed vertical line in Fig. \ref{fig:characterization}(c). A more dramatic change in slope is shown clearly in the evolution of b/a with temperature, where an inflection point is reached at \TN\: (inset in Fig. \ref{fig:characterization}(c)). 

The diffraction data of \ch{CoPS_3} point to the magnetostriction effect, where the onset of long-range magnetic ordering causes changes in the lattice parameters, reflected in the lattice expansion and contraction. Further evidence for this structural component to the phase transition at \TN\:  is given by an examination of the specific heat capacity data of the magnetic transition (Figure \ref{fig:characterization}(d)). Here we see a peak in the specific heat capacity around \TN, with the $\Delta$\textit{C$_P$} at the transition  is $\approx$ 36 J/(mol K). For comparison, in \ch{FePS_3}, the antiferromagnetic transition is accompanied by a structural transition, leading to a $\Delta$\textit{C$_P$} of the same order of magnitude as our measurement ($\approx$ 65 J/(mol K))\cite{TAKANO2004}.  In contrast, $\Delta$\textit{C$_P$} in \ch{MnPS_3} was reported to be $<$1 J/(mol K) at its magnetic transition temperature\cite{TAKANO2004}, where there is no evidence for a structural component to the magnetic transition. We also performed heat capacity measurments on \ch{ZnPS_3}, which is a paramagnetic sister compound to \ch{CoPS_3}; this material does not exhibit any peaks across the same temperature range. A recent report on \ch{CoPS_3} \cite{Koughnet2024}, using an IR reflection over a wide spectral range, showed that the Co$^{2+}$ cation site symmetry is lowered when proceeding into the AFM state as the number of electronic absorption bands increased from 4 to 6; these data were correlated with a change in the temperature-dependence of the layer spacing at \TN. These observations are suggestive of a subtle structural transition occurring at \TN\:involving the Co$^{2+}$ cation. 

\begin{figure*}[]
\includegraphics[width=0.7\textwidth]{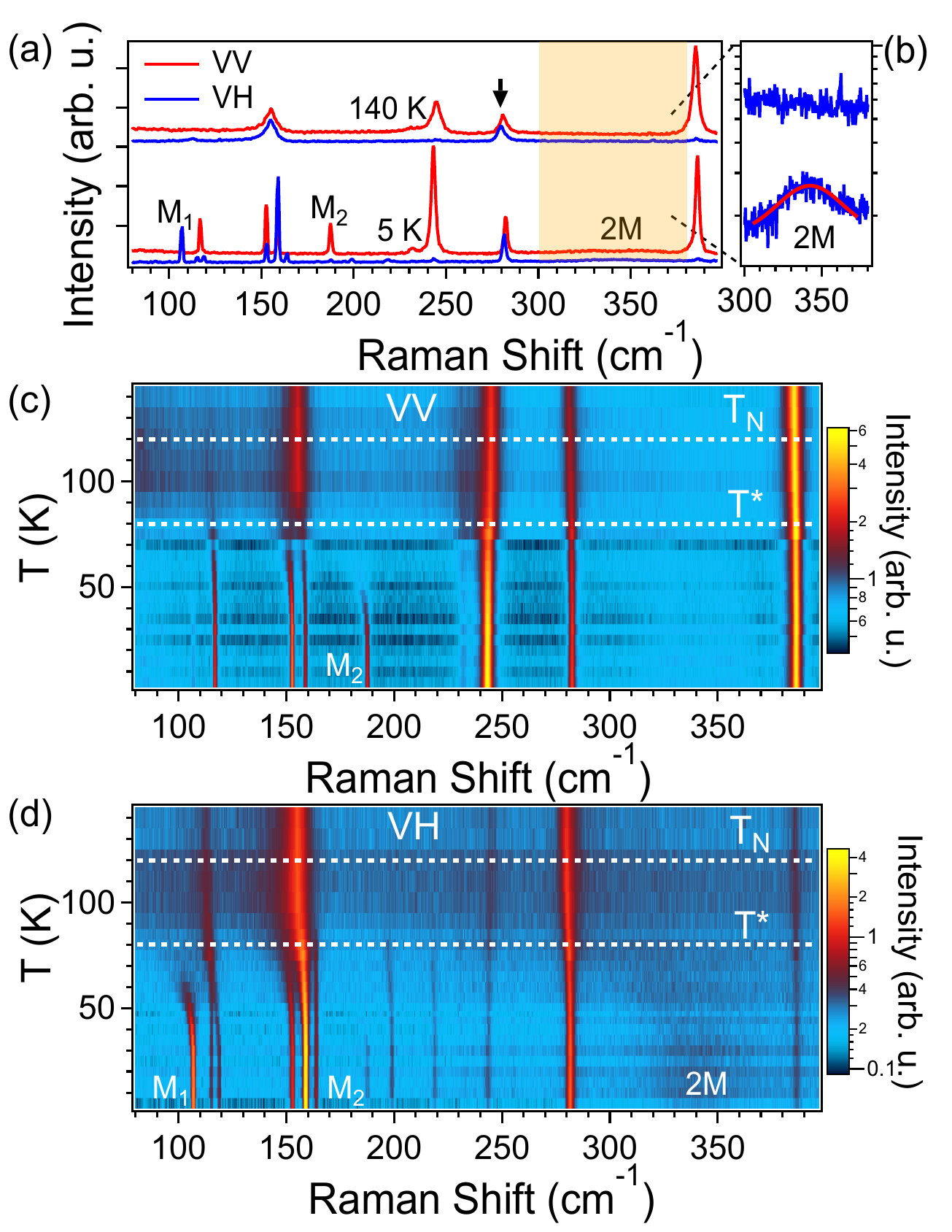}
\caption{\textbf{Temperature-dependent Raman spectra from CoPS$_3$}. (a) High (140 K) and low (5 K) Raman spectra of \ch{CoPS_3} are shown for parallel (VV) and crossed (VH) polarization, with offsets for clarity. The arrow points to a pair of \Ag{}/\Bg{} phonons with nearly degenerate frequencies, \textit{above and below} \TN. Zone-center magnon modes are marked \M{1} and \M{2}. The 2-magnon scattering peak is labled as 2M in the highlighted region between 300 \wn to 380 \wn. (b) A zoomed in plot onto the 2M region, of the VH spectra at 140 K (above) and 5 K (below), with the intensity on a log scale. A Voigt lineshape is fitted to the 5 K spectrum, with center frequency 342.6 \wn. False color plot of the Raman spectra as a function of temperature for VV (c) and VH (d), with a log intensity scale. Dash lines represent \TN\:and $T^*$, where a number of new modes appear. }
\label{fig:Tdep}
\end{figure*}

Next, we discuss temperature dependent Raman scattering across the phase transition. We observed drastic changes between the Raman spectra taken at 140 K and 5 K, above and below \TN\:  (Fig. \ref{fig:Tdep}(a)).  The measured peak frequencies at 5 K are tabulated in Table \ref{tab:RamanModes}. Two very obvious differences between the high and low temperature spectra are: the sharpness of the low temperature peaks, and the appearance of new modes. We labeled two new modes at low temperatures as \M{1} and \M{2}, as they have the same energy as the zone-center magnons measured by inelastic neutron scattering \cite{wildes_CoPS3_2023}.

Subtle differences in some phonon frequencies below and even above \TN\:between co- and cross-polarization configurations (VV and VH, respectively), can be seen in \Ph{11} and \Ph{12} (black arrow in Fig. \ref{fig:Tdep}(a)). This behavior is the hallmark of a high-quality single crystal, as seen in other 2D materials belonging to the same space group C2/m (point group C$_{\rm{2h}}$)\cite{Larson_CrI3_2018,Mai_RuCl3_2019}. The small energy splitting can be understood as the symmetry breaking effect of the 3-fold rotational symmetry of each individual layer due to the monoclinic stacking, with the energy scale proportional to the vdW interaction\cite{Larson_CrI3_2018}. In these weakly coupled vdW layers, such monoclinic stacking can be destroyed by handling of the sample \cite{Cao_RuCl3_2016}. This would manifest as a single broad phonon peak in all polarization configurations. The broadened Raman peak can lead to the interpretation that the bulk crystal exhibits 3-fold rotational symmetry of the single vdW layer, of point group D$_{\rm{3d}}$.

\renewcommand{\arraystretch}{1.2}
\begin{table}[b]
\caption{\textbf{Selected Raman peak frequencies of \ch{CoPS_3} at 5\,K}. The experimental Raman peak frequencies at 5\,K from fitting the spectra using a Voigt line shape compared with calculated values from density functional theory (DFT) for Raman active phonons. We assign the DFT calculated phonon frequencies (and their symmetries) to the experimental value based on its presence in the paramagnetic phase. Zone-folded (ZF) modes calculated frequency and symmetry assignment are intentionally left out due to the multiple number of possible bands that could match the experimental values. 
* The zone-center magnon modes, \M{1} and \M{2} have their symmetry assignment based on Fig. \ref{fig:Hybridization}(a) and Fig. S2.}
\begin{tabular}{|W{c}{45pt}|W{c}{45pt}|W{c}{68pt}|W{c}{68pt}|}
\hline
Label      &  Symmetry      &  DFT (\wn)   &   Exp. (\wn) \\ \hline
\Ph{1}      &  ZF       &  -		  &   115.2	\\ \hline
\Ph{2}      &  \Bg{}       &  103.4		  &   116.9	\\ \hline
\Ph{3}      &  \Ag{}       &  104.3		  &   118.8	\\ \hline
\Ph{4}      &  \Bg{}       &  163.4		  &   152.6	\\ \hline
\Ph{5}      &  \Ag{}       &  163.9		  &   158.7 \\ \hline
\Ph{6}      &  ZF       &  -		  &   163.8	\\ \hline
\Ph{7}      &  ZF       &  -		  &   198.9	\\ \hline
\Ph{8}      &  ZF       &  -		  &   218.6	\\ \hline
\Ph{9}      &  \Ag{}       &  218.5		  &   231.8	\\ \hline
\Ph{10}      &  \Ag{}       &  233.8		  &   243.1	\\ \hline
\Ph{11}      &  \Bg{}       &  267.1    	  &   281.4	\\ \hline
\Ph{12}      &  \Ag{}       &  268.2		  &   282.3	\\ \hline
\Ph{13}      &  \Ag{}       &  369.1		  &   386.1	\\ \hline
\M{1}      &  \Ag{}*       &  	-	  &   106.9	\\ \hline
\M{2}      &  \Bg{}*       &  	-	  &   187.5	\\ \hline
\end{tabular}
\label{tab:RamanModes}
\end{table}

The most dramatic changes can be seen in the frequency region below 220 \wn. There are several new peaks appearing, in the VV and VH configurations (Fig. \ref{fig:Tdep}(c) and (d)), while two clusters of peaks around 115 \wn and 155 \wn, are seemingly formed by the spliting of two significantly broader peaks at high temperature. This behavior is strikingly similar to previous reports on \ch{FePS_3} \cite{Balkanski_MPX3_1987,Scagliotti_FePX3_1987,lee_FePS3_2016,McCreary_FePS3_2020}, as well as \ch{FePSe_3} \cite{cui_FePSe3_2023,luo_FePSe3_2023}, where two broad Raman scattering peaks above \TN\:split into multiple sharp Raman modes below \TN. Curiously, the temperature at which these phenomena occur in \ch{CoPS_3} is well below \TN\:$\approx$ 119\,K, which we denote as $T^* \sim$ 80\,K. It should be noted that in this temperature range, the lattice anharmonic effect is inconsequential. In contrast, \ch{FePS_3} with \TN\:$\approx$ 118\,K, the new Raman modes appeared only $\sim$ 8 K below \TN, at $\approx$ 110\,K. In \ch{CoPS_3}, we observed a subtle jump in the heat capacity near $T^*$, possibly connected to the observations with Raman scattering (SI section IV).

\begin{figure*}[]
\includegraphics[width=.7\textwidth]{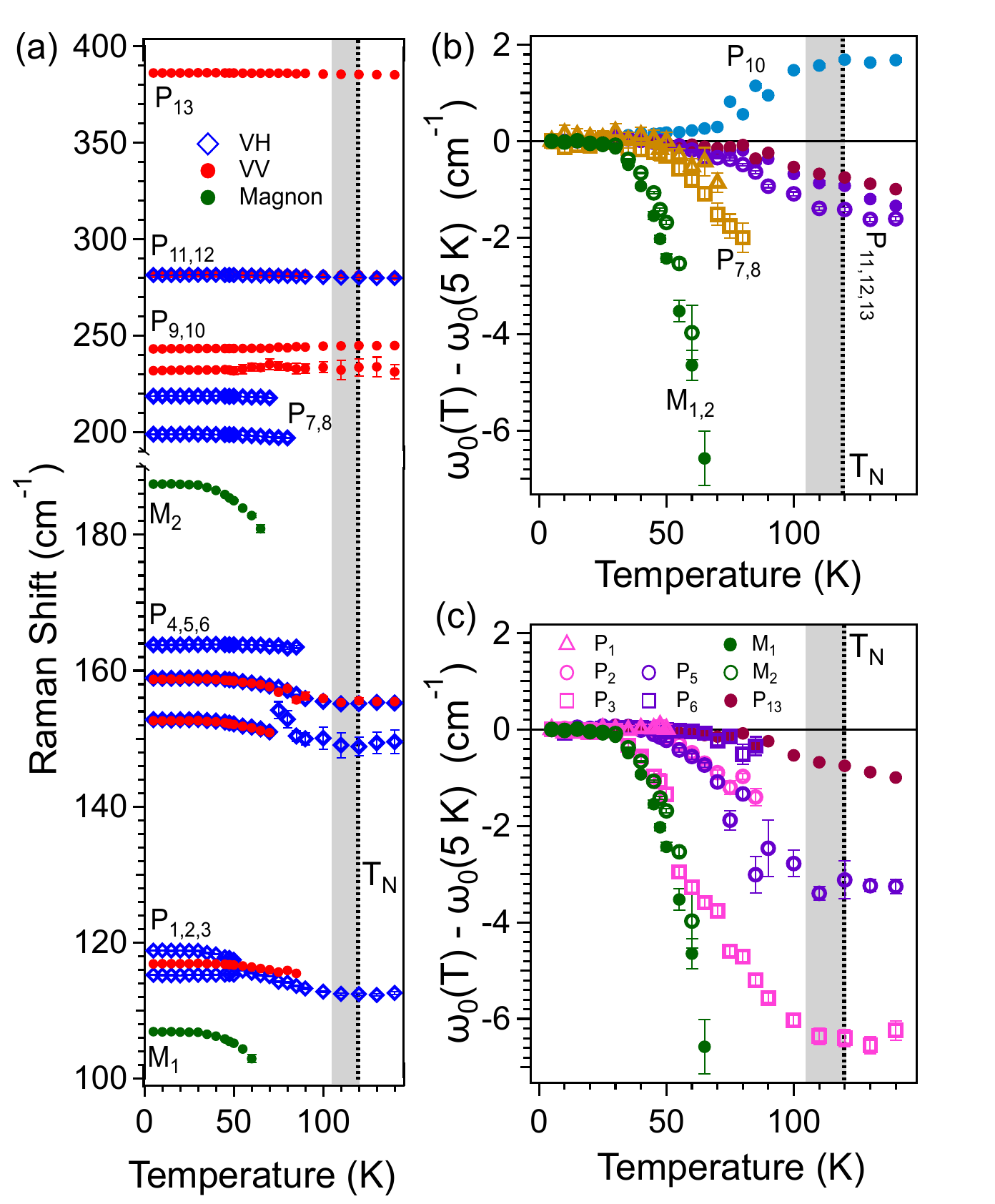}
\caption{\textbf{Temperature-dependent Raman peak frequencies}. (a) Peak frequencies of Raman modes in both VV and VH polarization configurations, labled as \Ph{1,..,13}. The two magnons are once again labeled as \M{1,2}. (b) Change in peak frequency as a function of temperature of select modes, showing anomalous (\Ph{10}), "typical" \Ph{11,12,13}, zone-folding modes \Ph{7,8}, as well as the two magnons \M{1,2}. (c) The same change in Raman peak frequency for two peak clusters around 116 \wn and 157 \wn, where evidence of hybridization of \M{1} and \Ph{3} can be seen by the strong frequency shift of \Ph{3} (hollowed magenta squares) below 50 K, where lattice effects are minimal (as seen in other phonons). The error bars represent 1 std. dev. from the fitting procedure. The N\'eel transition as well as the critical spin fluctuations \cite{wildes_CoPS3_2017} are marked by the vertical dash lines and shaded regions respectively in each plot.}
\label{fig:TdepFits}
\end{figure*}

We fit the Raman spectra using the Voigt lineshape; the fitted peak frequencies exhibit a complex temperature dependence visible in Fig. \ref{fig:TdepFits}(a). The peaks in Fig. \ref{fig:TdepFits}(a) are labeled corresponding to Table 1. In Figure \ref{fig:TdepFits}(b) and \ref{fig:TdepFits}(c), two Brillouin-zone-centered magnon modes, \M{1} and \M{2}, can be seen having the strongest red-shift, more than 6 \wn, as we warm up towards \TN. This behavior is typical of a one-magnon excitation where thermal fluctuations renormalize the magnon energy\cite{Lockwood_1984,McCreary_FePS3_2020}. We also found a broad and weak feature at 343 \wn ($\approx$ 42 meV) at 5 K (Fig. \ref{fig:Tdep}(b), (d)). This is consistent with a 2-magnon (2M) excitation, similar to the one found in \ch{NiPS_3} \cite{Rosenblum_NiPS3_1994,kim_NiPS3_2019}.

Interestingly, the broad 2M scattering peak at 343 \wn evolves into a scattering continuum across our measured spectral range above $T^*$ (continuous dark color background in Fig. \ref{fig:Tdep}(d), more clearly in Fig. S4). Drastic changes in the Raman spectrum occur around the same temperature, as mentioned above. The broader spectral linewidth above $T^*$ means a shorter lifetime for the phonon. Such shortened phonon lifetime can be caused by an additional decay pathway, i.e. into the broad magnetic excitation continuum. The correlation between the magnetic scattering continuum and the broad Raman phonon peaks points to significant coupling of the lattice degrees of freedom, represented by the phonons, to the magnetism (see the comparison between \Ph{5} linewidth and scattering background temperature dependent in Fig. S4). 

Our analysis of the Raman peak frequencies also reveals at least 4 other modes that disappear with increasing temperature, \Ph{1}, \Ph{6}, \Ph{7}, \Ph{8}. All of these modes are not measurable above $T^*$. We attribute these disappearances to the effect of zone-folding. The magnetic ordering vector was measured to be [0 1 0] by neutron scattering \cite{wildes_CoPS3_2017}. Since the unit cell of space group C2/m is twice that of the primitive cell in the ab-plane, the propagating vector of [0 1 0] implies that the periodicity of the magnetic structure is twice that of the primitive cell size in the b-direction. Thus, the magnetic structure provides the necessary crystallographic momentum for zone-folding to occur, i.e. Umklapp scattering. The density functional theory (DFT) calculated normal and zone-folded peaks are included in Table \ref{tab:RamanModes}. Interestingly, we found that the linewidths of the new modes are extremely sharp at low temperature, beyond our instrument’s limit of 1.5 \wn \cite{Tuschel_2020}. 

Figure \ref{fig:TdepFits} (b) and (c) shows the difference in frequency of the observed modes to their corresponding frequencies at 5 K as a function of temperature. The N\'eel transition is marked at 119 K (dash line), as well as the critical fluctuation region down to 109 K \cite{wildes_CoPS3_2017} (shaded region). In this temperature range, below 140 K, \Ph{11,12,13} can be seen exhibiting slight anharmonic frequency softening as temperature increases to \TN\: ( Fig. \ref{fig:TdepFits}(b)), with changes in phonon frequency around 1 \wn. \Ph{10} shows an anomalous temperature dependent behavior, with increasing frequency as temperature is increasing. We suspect the renormalization of this phonon frequency is tied to its modulation of the exchange interaction, similar to what has been measured in magnetoelectric materials \cite{Sushkov_spinphonon_2005,Park_CoBO_2023}.
Interestingly, there is a kink in the temperature dependent frequency of \Ph{10} as well as \Ph{11,12,13}, around 70 K to 80 K (Fig. \ref{fig:TdepFits}(b)). Some of the zone-folded modes, \Ph{7} and \Ph{8}, show a slightly stronger temperature dependence than the typical phonons (e.g. \Ph{11}). It is reasonable to assume that the zone-edge scattering wavevector that allows for the observation of these modes with Raman scattering would be impacted by the magnetic fluctuations near the phase transition. 

The fitting results reveal a complex story for the two clusters of modes, \Ph{1,2,3} and \Ph{4,5,6}. As seen in Fig. \ref{fig:TdepFits}(a), we assign \Ph{2} and \Ph{6} as the zone-folded modes by their sudden disappearance around $T^*$. While all of these modes show some frequency changes (Fig. \ref{fig:TdepFits}(c)), the most notable changes occur at temperatures well below \TN.

\begin{figure*}[]
\includegraphics[width=.8\textwidth]{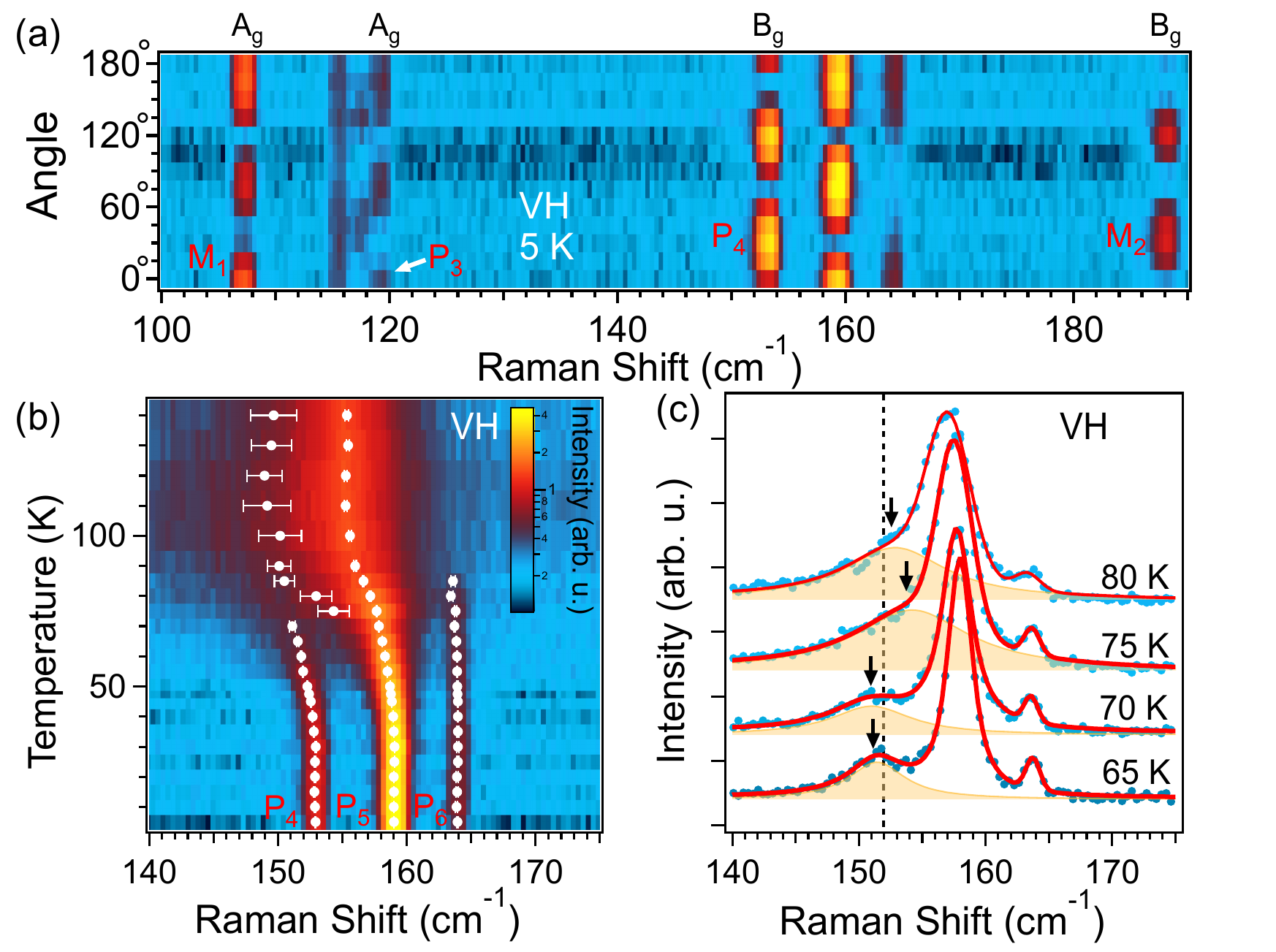}
\caption{\textbf{Symmetry and hybridization between magnons and phonons in CoPS$_3$}. (a) False color polot of polarization-resolved Raman spectra at 5 K, in VH configuration, focusing on two phonon clusters around 116 \wn and 157 \wn. The irreducible representation labels of the 2/m point group are included to show pairs of modes with the same symmetry: \M{1} and \Ph{3} (\Ag{}), \M{2} and \Ph{4} (\Bg{}). (b) False color plot of of the temperature dependent Raman spectra in VH configuration focused on the cluster at 157 \wn, with peak fitting results overlayed to show the avoided crossing phenomenon between 70 K and 75 K. The log-scale intensities are the same for (a) and (b). (c) Select spectra around 75 K in VH configuration with fitted results of three Voigt-lineshape peaks (solid lines) and specifically \Ph{4} (shaded peak with arrows representing center frequencies). The vertical dash line highlights the sudden jump in the center frequency of \Ph{4}.}
\label{fig:Hybridization}
\end{figure*}

The first cluster around 116 \wn undergoes an apparent “splitting” around 50 K. At high temperature, we can only fit a single peak. We expect there to be a pair of phonons with \Ag{} and \Bg{} symmetry at this frequency. This lack of observation of the second peak is likely due to its weaker signal and significant linewidth broadening. We subtract the high temperature frequency from the mode at low temperature that produces the least discontinuity, which turns out to be \Ph{3}. Below 50 K, we see a strong temperature dependence of \Ph{3}’s frequency, comparable to the magnons (Fig. \ref{fig:TdepFits}(c)). This is the signature of mode repulsion, since at 50 K the phonon frequency should be frozen. Such an observation is evidence of hybridization between \Ph{3} and \M{1}. Below 50 K, we see almost identical temperature dependence between \M{1} and \Ph{3}, where significant change ($> 2$\wn) is followed by a flattening out around 30 K. Similar behavior was observed in \ch{FePSe_3}, where the magnon and phonon are hybridized at low temperature \cite{cui_FePSe3_2023,luo_FePSe3_2023}. The likely hybridization between \M{1} and \Ph{3} is further supported by the fact that they share the same polarization dependence, and hence the same symmetry, at 5 K (Fig. \ref{fig:Hybridization}(a)). 

Phenomenologically, the interaction and subsequent mixing between two quantum states can occur if their wavefunctions overlap. The irreducible representations (IRREPs) of the 2/m point group, e.g. \Ag{} or \Bg{}, provide a simple way to confirm the possible overlap or orthogonality of these wavefunctions. Fig. \ref{fig:Hybridization}(a) shows that \M{1} (\M{2}) exhibits a similar polarization dependence to an \Ag{} (\Bg{}) phonon (see SI section III for more details). This implies that the wavefunction of \M{1} transforms like the IRREP \Ag{}, which is orthogonal to \Bg{}. Thus, \M{1} is allowed to interact with an \Ag{} phonon (\Ph{3}) and not a \Bg{} phonon (\Ph{2}). Similarly, \M{2} is allowed to interact with a \Bg{} phonon (\Ph{4}) and not an \Ag{} phonon (\Ph{5}). Indeed, we also discovered the interaction between \M{2} and \Ph{4} (and \textit{not} \Ph{5}).

In the second cluster around 157 \wn (Fig. \ref{fig:Hybridization} (b) and (c)), we observed two broad peaks at high temperature. This is predicted from DFT, as we expected a pair of \Ag{} and \Bg{} phonons, \Ph{5} and \Ph{4}. At 70 K, the fitting results reveal an avoided crossing behavior in \Ph{4}, just before \M{2} exhibits measurable intensity at much higher frequency than \Ph{4}. The fitting results and false color spectra are presented in Fig. \ref{fig:Hybridization} (b), as well as several spectra around the avoided crossing superimposed with individual fitted peaks in Fig. \ref{fig:Hybridization}(c). These observations indicate that as the magnon frequency changes with temperature and begins to overlap with \Ph{4} energetically, an avoided crossing occurs. We estimate the splitting energy to be 3 \wn, or 0.4 meV. The magnon-phonon coupling strength is estimated to be $\approx$ 3\wn \cite{mag-phonon-estimation}, comparable to previously observed values of the coupling energy in 3d transition metal ferro- and ferrimagnetic compounds\cite{Kang_2010,Brinzari_2013}.

\begin{figure*}[]
\includegraphics[width=.9\textwidth]{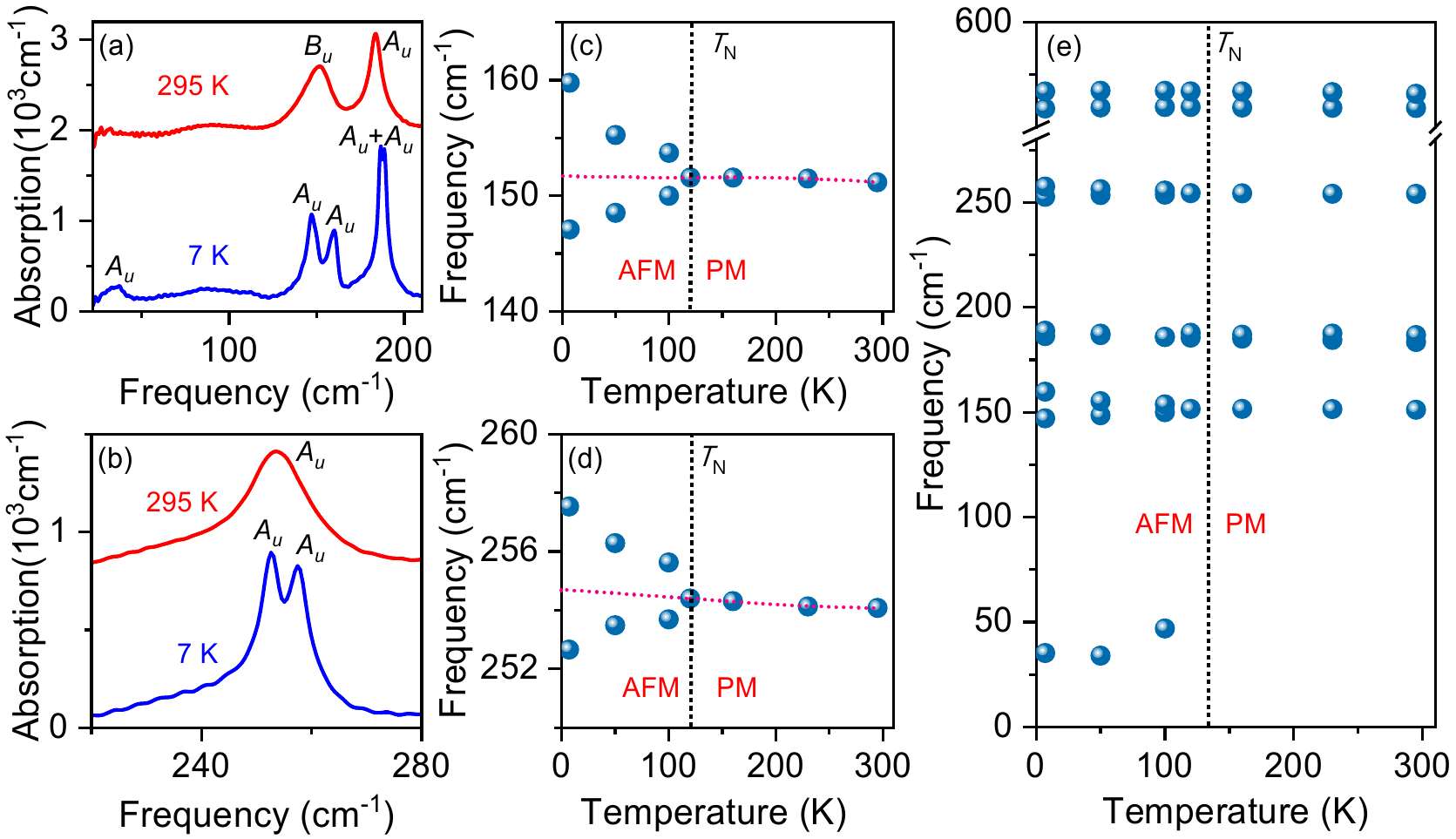}
\caption{\textbf{Infrared response of CoPS$_3$}. (a,b) Close-up view of the infrared absorption in the high and low temperature phases along with (c,d) frequency vs. temperature plots showing spectral changes across the 119 K N\'eel transition. In panels (c,d), an anharmonic model is fit to the high temperature phase data. (e) Full view of the frequency vs. temperature trends.}
\label{fig:Infrared}
\end{figure*}

The infrared absorption data of CoPS$_3$ further supports these findings. Here, evidence for spin-phonon coupling is present in the form of peak shifting and splitting of \Au{} and \Bu{} modes across the 119 K magnetic ordering transition (Fig. \ref{fig:Infrared}). Extracting the frequency shifts of these branches from an anharmonic fit of the high temperature phase data and assuming that the spin-spin correlation function ${\langle}S_i{\cdot}S_j{\rangle}$ goes as $S^2$ = (3/2)$^2$ = 9/4 in the low temperature limit, we can estimate the spin-phonon coupling constants ($\lambda$'s) for the different  branches of these modes as ${\Delta}{\omega} = {\lambda}{\langle}S_i{\cdot}S_j{\rangle}$ \cite{Park_CoBO_2023}. We find ${\lambda}$'s for the two branches of the 151 cm$^{-1}$ $A_u$ symmetry mode below $T_{\rm N}$  are approximately - 2.02 and + 3.6 cm$^{-1}$. The ${\lambda}$'s for the two branches of the 252 cm$^{-1}$ \Bu{} symmetry mode  to be on the order of $\pm$ 1 cm$^{-1}$, respectively. These values are an order of magnitude smaller than what is observed in related Co-containing oxides \cite{Park_CoBO_2023}. 

A recent study \cite{Rao_2024} showed a similar deviation from the phonon anharmonicity effects, for inversion-even Raman phonons, due to a general spin-phonon coupling effect, down to 100 K. It should be noted that only the two clusters mentioned above undergo striking changes in linewidth and splitting below $T^*$. In comparison, the phonons at higher energy remain single peaks with little changes to their frequency and lifetime across \TN. Non-linear optical spectroscopy measurements have been carried out recently for \ch{FePS_3}\cite{Ergecen_2023} and \ch{CoPS_3}\cite{Khusyainov_2023} that reveal coherent spin-lattice coupling in these materials. Specifically, the strong coupling between the spin-orbital electronic excitation of the Co$^{2+}$ to phonon modes at 3.5 THz ($\approx$ 117 \wn) and 4.7 THz ($\approx$ 157 \wn) seen by \citep{Khusyainov_2023} further corroborate our finding of the entanglement effect between the spin and lattice in \ch{CoPS_3}.

\section{Discussion}

Previously, there has been no clear evidence of a second phase transition or crossover behavior in \ch{CoPS_3} below \TN\:from inelastic neutron scattering and magnetic susceptibility\cite{wildes_CoPS3_2017}. In the present study, the temperature dependent phonon spectra exhibit splitting of Raman phonon modes, at some intermediate temperature $T^* \sim 80$ K below \TN. Furthermore, we detect no anomaly in the heat capacity between 30 K and 100 K (Fig. S3). Combined with the lack of a clear lattice symmetry breaking signature in the high-resolution X-ray data at such a temperature, the data point to a more subtle effect that occurs near $T^*$. A recent publication on few layers \ch{NiPS_3} \cite{Sun_2024} linked the splitting of the broad Raman phonons as well as the appearance of new peaks at low temperature to the broken 3-fold rotational symmetry and broken translational symmetry (zone-folding) of the individual vdW layer, respectively, caused by the spontaneous long range magnetic ordering. While in the few layers limit, rotational symmetry breaking effect of the monoclinic stacking should become less significant, we have no reason to believe that the high temperature Raman and IR spectra from our bulk sample to be from the individual layer responses.

Indeed, we already observed the effect of broken 3-fold rotational symmetry above \TN\: (Fig. \ref{fig:Tdep}(a)). Due to this observation, the dramatic differences between the phonon spectrum above and below $T^*$ remain a challenging puzzle to address. In this work, we presented a qualitative picture of spin-lattice coupling based on zone-folding, magnon-phonon hybridization, and a broad magnetic scattering continuum that exists only above $T^*$. The data indicate the coherent coupling between the lattice and magnetic normal modes at low temperature, embodied by the avoided crossing around 75 K (Fig. \ref{fig:Hybridization}. The microscopic details of these interactions between the lattice and magnetic ground state excitations remain to be studied. Calculation of the magnon-phonon coupling could be carried out in the framework of phonon modulation of exchange pathway \cite{Sushkov_spinphonon_2005}. Nevertheless, such unique entanglement between specific crystal lattice and spin lattice excitations would imply that at some higher temperature, thermal spin fluctuations can lead to the broadening linewidth of specific phonon modes.

The various temperature dependent behaviors described here are not unique to \ch{CoPS_3}. Some can be found in other magnetic 3d transition metal members of the \ch{MPX_3} family. However, only in \ch{CoPS_3} are these spin-lattice effects found together and with significant magnitude, possibly due to the coupled spin and orbital moments in Co$^{2+}$. Our observations both highlight the complexity of coupled degrees of freedom in real-world quasi-two-dimensional materials, as well as showcasing the \ch{MPX_3} family as the perfect platform to study these couplings. 

\section{Summary} 

In summary, we present a detailed experimental study of the spin and lattice coupling in \ch{CoPS_3}. The bulk characterization results point to a concomitant structural anomaly with the antiferromagnetic transition, indicative of magnetoelastic coupling. While such behavior is typical in the magnetic members of the \ch{MPX_3} family \cite{Bjarman1983,JERNBERG1984,TAKANO2004}, we also observed abrupt changes to the phonon spectrum around $T^* \sim 80$ K, below \TN\:$\approx 119$ K. Among the changes is the appearance of new modes, some of which are Brillouin zone center magnons \cite{wildes_CoPS3_2023}, and zone-folded phonons due to the magnetism-induced doubling of the unit cell. The latter is confirmed by first-principles calculation.
We also discovered the coherent coupling between two pairs of magnons and phonons, at approximately 116 \wn and 153 \wn. This observation is a tell-tale sign of the strong entanglement between the crystal lattice and the magnetism in \ch{CoPS_3}. We employed the mode repulsion phenomenology to quantify the observed magnon-phonon coupling effect. Furthermore, we point out the need for a microscopic model that can capture the coupling between the magnetic excitation spectrum and the lattice excitation spectrum in \ch{CoPS_3}, in order to elucidate the coupling effect between spin and lattice in these 2D magnetic materials.

\section*{Acknowledgements}
Work at Tennessee (SS and JLM) is supported by Physical Behavior of Materials, Basic Energy Sciences, U.S. Department of Energy (Contract number DE-SC0023144).

Work at University of Virginia (MZ and MNAT) is supported by the National Science Foundation under Grant No. 2421213.

Use of the Advanced Photon Source at Argonne National Laboratory was supported by the U. S. Department of Energy, Office of Science, Office of Basic Energy Sciences, under Contract No. DE-AC02-06CH11357.

Work at Oak Ridge National Laboratory was supported by the U.S. Department of Energy, Office of Science, Basic Energy Sciences, Materials Sciences and Engineering Division.

The work at AFRL is supported by the Air Force Office of Scientific Research (AFOSR) grant no. LRIR 23RXCOR003.

Certain trade names and company products are identified in order to specify adequately the experimental procedure. In no case does such identification imply recommendation or endorsement by the National Institute of Standards and Technology, nor does it imply that the products are necessarily the best for the purpose.

DISTRIBUTION STATEMENT A. Approved for public release: distribution unlimited. AFRL-2025-2706

\bibliography{CoPS3.bib}

\begin{thebibliography}{50}
\bibitem [{mag()}]{mag-phonon-estimation}%
  \BibitemOpen
  \href@noop {} {}\bibinfo {note} {We estimate the splitting and coupling constant based on the energy deflection of the measured phonon peak. In a coupled 2-state system, the maximum splitting is twice that of the coupling constant. Hence, the maximum deflection measured in a single mode is approximately 1/2 of the splitting, which is the coupling constant.}\BibitemShut {Stop}%
\end{thebibliography}

\end{document}